\documentclass[openacc]{rstransa}%%%%where rstrans is the template name

\usepackage{slashed}
\usepackage{amsmath}
\usepackage{amsfonts}
\usepackage{amssymb}

\newcommand{\hL}{h_\Lambda}

\newcommand{\mL}{m_\Lambda^2}
\newcommand{\lL}{\lambda_{2,\Lambda}}
\newcommand{\mH}{m_{\text{H}}}
\newcommand{\mtop}{m_{\text{t}}}

\newcommand{\pat}{\partial_t}
\newcommand{\Tr}{\text{Tr}\,}
\newcommand{\Eqref}[1]{Eq.~\eqref{#1}}
\newcommand{\Uss}{U_{\text{eff}}}

\begin{document}

%%%% Article title to be placed here
\title{Renormalization Group Flow of the Higgs Potential}

\author{%%%% Author details
Holger Gies$^{1,2,3}$ and Ren\'{e} Sondenheimer$^{1}$}

%%%%%%%%% Insert author address here
\address{$^{1}$Theoretisch-Physikalisches Institut, %
Friedrich-Schiller-Universit\"at Jena, Max-Wien-Platz 1, 07743 Jena, Germany\\
$^{2}$Abbe Center of Photonics, Friedrich Schiller University Jena, Max Wien 
Platz 1, 07743 Jena, Germany\\
$^{3}$Helmholtz-Institut Jena, Fr\"obelstieg 3, D-07743 Jena, Germany}

%%%% Subject entries to be placed here %%%%
\subject{xxxx, xxxx, xxxx}

%%%% Keyword entries to be placed here %%%%
\keywords{Standard Model, Higgs Boson, Renormalization Group}

%%%% Insert corresponding author and its email address}
\corres{Holger Gies\\
\email{holger.gies@uni-jena.de}}

%%%% Abstract text to be placed here %%%%%%%%%%%%
\begin{abstract}
  We summarize results for local and global properties of the
  effective potential for the Higgs boson obtained from the functional
  renormalization group, which allows to describe the effective
  potential as a function of both scalar field amplitude and RG
  scale. This sheds light onto the limitations of standard estimates
  which rely on the identification of the two scales and helps
  clarifying the origin of a possible property of meta-stability of
  the Higgs potential. We demonstrate that the inclusion of
  higher-dimensional operators induced by an underlying theory at a
  high scale (GUT or Planck scale) can relax the conventional lower
  bound on the Higgs mass derived from the criterion of absolute
  stability.
\end{abstract}
%%%%%%%%%%%%%%%%%%%%%%%%%%%

%%%%%%%%%% Insert the texts which can accomdate on firstpage in the tag "fmtext" %%%%%

\begin{fmtext}
\section{Introduction}
%%%% Insert A head here

The measurement of the mass of the Higgs boson
\cite{Aad:2012tfa,Chatrchyan:2012xdj} together with masses and
couplings of the other standard model degrees of freedom has made it
clear that the standard model happens to reside in a rather particular
place in parameter space: extrapolating the renormalization running of
the coulplings to higher scales, all couplings (apart from the U(1)
gauge coupling) tend to smaller values, possibly towards zero. The
Higgs-field self coupling may even drop below zero, which is
conventionally interpreted as a signature for a potential instability
of the standard model occuring at a scale near $10^{10\dots12}$
GeV.

\end{fmtext}

%%%%%%%%%%%%%%% End of first page %%%%%%%%%%%%%%%%%%%%%

\maketitle

Since the standard model as a quantum field theory is conceptually
defined in terms of a functional integral and a bare action that
parametrizes the microscopic interactions, it is useful to look at the
system from a top-down perspective starting from a high energy scale
$\Lambda$, where the microscopic interactions are fixed.  The
long-range observables measured at colliders then are a result
obtained after averaging over the fluctuations of all quantum degrees
of freedom. Depending on the embedding into an underlying theory,
$\Lambda$ may be considered as a GUT-like scale or the Planck scale,
or any other scale where degrees of freedom beyond the standard model
contribute significantly to the dynamics. Technically, $\Lambda$ may be viewed as a
UV cutoff regularizing the highest energy scales where the standard
model description does no longer apply anyway.

From this viewpoint, all long-range observables are fixed, once the
bare action $S_{\Lambda}$ is chosen. A convenient tool to bridge the
gap between $\Lambda$ and collider scales is the renormalization group
(RG), quantifying the running of couplings and masses from the UV to
the IR. Even in the weak coupling regime, the dependence of the IR
observables on the bare parameters in $S_{\Lambda}$ can be involved
and requires the solution of the RG flow. Even before the first
measurement of the Higgs mass, it has long been known that possible
mass values are bounded by a finite interval, the IR window
\cite{Maiani:1977cg,Krasnikov:1978pu,Lindner:1985uk,Wetterich:1987az,Altarelli:1994rb,Schrempp:1996fb,Hambye:1996wb},
once a set of parameters are fixed, most prominently the heaviest top quark mass. 

The measured value of the Higgs mass appears to indicate that $\mH$ is
slightly below the lower bound imposed by demanding for absolute
stability of the electroweak Fermi vacuum, while it is well inside the
bound, if the Fermi minimum is permitted to be meta-stable but
sufficiently long lived, see, e.g.,
\cite{Buttazzo:2013uya,Espinosa:2015kwx}. While a precise location of
the bound still requires a better accuracy for the determination of
the top Yukawa coupling and the strong coupling constant
\cite{Bezrukov:2014ina}, we wish to emphasize and quantify the
influence of the bare action $S_{\Lambda}$ that comes along with a
large number of unknown and not directly measurable parameters. From
the RG perspective, most of these parameters are RG irrelevant.

While the perturbative RG typically concentrates on the RG relevant
operators, presupposing perturbative renormalizability, the functional
RG can also account for power-counting irrelevant operators. As long
as the UV scale $\Lambda$ is finite, also the irrelevant operators can
contribute to long-range observables and thus to the size of the IR
window. By power-counting arguments, their influence on long-range
observables is typically powerlaw suppressed. Nevertheless, we argue
below that they do have a quantitative impact, e.g., on the precise
location of the line separating the fully stable from the meta-stable
case. In this sense, also the irrelevant operators can be essential
when drawing conclusions about the fate of the universe as we know it.

\section{Perturbative RG vs. fermion determinant}

In order to keep the discussion simple, we use a toy model for the
top-Higgs sector involving a Dirac fermion and a real scalar field,
featuring a discrete chiral $\mathbb{Z}_2$ symmetry \cite{Gies:2013fua}. We
parametrize the bare action as
\begin{equation}
 S_\Lambda = \int d^4x \left[ \frac{1}{2} (\partial_\mu \phi)^2 + U_{\Lambda}(\phi) + \bar{\psi}i\slashed{\partial}\psi + i{\hL} \, \phi\bar{\psi}\psi \right],
 \label{eq:action}
\end{equation}
with a bare Yukawa coupling ${\hL}$ and a bare potential
$U_{\Lambda}$. This simple model shares many features relevant for the
stability problem with the standard model. We comment on more
extensive models and the standard model below.

A conventional simple estimate of the effective potential of the Higgs
is given by using a Coleman-Weinberg inspired form of the potential,
\begin{equation}
\Uss(\phi)= -\frac{1}{2} \mu^2 \phi^2 + \frac{1}{8} \lambda(\phi) \phi^4, 
\label{eq:Usinglescale}
\end{equation}
where the mass parameter $\mu$ is chosen such that the potential
acquires a vacuum expectation value $v$ at the Fermi scale $v\simeq
246$GeV. Here, $\lambda(\phi)$ is determined from RG-improved
perturbation theory, i.e., by integrating the $\beta$ function of the
coupling which reads to one-loop order
\begin{equation}
\pat\lambda = \frac{1}{16\pi^2}\big(9\lambda^2 + 8h^2 \lambda - 16 h^4\big), \quad \partial_t=k\frac{d}{dk}. 
\label{eq:betalambda1l}
\end{equation}
The integration of \Eqref{eq:betalambda1l} results in the coupling
being a function of the RG scale $k$. The effective coupling for the
potential is then obtained by identifying the RG scale with the field
amplitude $\lambda(k=\phi)$. While this procedure is quantitatively
well justified in many cases (in particular in the absence of further
relevant scales), it comes with a loss of information: as a matter of
principle, the effective potential depends on both scales, the field
amplitude $\phi$ and the RG scale $k$ separately. In contrast to the
estimate for the ``single-scale'' potential \eqref{eq:Usinglescale}, we
can keep track of both dependencies with the functional RG.

The insufficiency of the single-scale potential becomes obvious from
the following puzzle: assuming that the top-Yukawa coupling $h$ at
some scale starts to dominate the flow towards the UV in
\Eqref{eq:betalambda1l}, the flow of $\lambda$ will decrease and
can even drop below zero. Upon insertion into $\Uss$, this can result in an
instability of the single-scale potential. This is the standard
argument for the occurrence of an instability also in the standard model.

Alternatively, we may not remain on the simple level of
\Eqref{eq:betalambda1l}, but compute the full top-quark contribution
to the effective potential to one-loop order. This is given by the fermion determinant,
\begin{equation}
U_{\text{F}}= - \frac{1}{2\Omega} \ln \frac{\det_\Lambda ( - \partial^2 + h^2 \phi^2)}{\det_\Lambda ( - \partial^2 )},
\end{equation}
where $\Omega$ denotes the spacetime volume, and the subscript
$\Lambda$ should remind us of the fact that a regularization at the
cutoff scale is necessary. The determinant can be worked out
exactly. For instance, for simple momentum-cutoff regularization, we get \cite{Gies:2014xha}
\begin{equation}
U_{\text{F}}= - \frac{\Lambda^2}{8\pi^2} h^2\phi^2 + \frac{1}{16\pi^2}
\left[ h^4 \phi^4 \ln \left( 1+ \frac{\Lambda^2}{h^2 \phi^2} \right) + h^2 \phi^2 \Lambda^2
- \Lambda^4 \ln \left( 1+ \frac{ h^2\phi^2}{\Lambda^2} \right) \right].
\label{eq:MFU}
\end{equation}
We observe a negative mass term $\sim \phi^2$. This is expected and
ultimately absorbed in the renormalization of the mass-like parameter
that fixes the Fermi scale as the vacuum expectation value $v$ of the
potential. Most importantly, the complete remainder being the
interaction part of the potential is manifestly positive at any finite
field amplitude.

The obvious puzzle now is that the same cause, namely the fermionic
fluctuations, seems to lead to two different effects: the reasoning
based on the single-scale potential suggests an instability for large
Yukawa coupling, whereas the fermions contribute strictly positively
to the full interaction potential in \Eqref{eq:MFU}.

The puzzle is resolved by noting that the exact result for
$U_{\text{F}}$ also keeps track of the cutoff $\Lambda$ dependence
which remains invisible in the single-scale approach
\cite{Holland:2003jr,Holland:2004sd,Branchina:2005tu,Gies:2013fua,Gies:2014xha}. While
perturbative renormalizability seems to suggest that the cutoff can be
removed by renormalization conditions, it cannot be sent to infinity
in the standard model because of its triviality problem in the U(1)
sector. Hence, $\Lambda$ is a (place holder for a) physical scale
which should be kept track of.

It is instructive to try to reconcile the two contrary ends of the
puzzle. For this, we may naively expand the interaction part of the
potential \eqref{eq:MFU} in powers of $h^2\phi^2/\Lambda^2$ for large
$\Lambda$. Then we may absorb the $\log \Lambda$ divergence into a
renormalization of $\lambda$, introducing a renormalization scale
$\Lambda\to k$, and end up with
\begin{equation}
U_{\text{F}}|_{\phi^4}\stackrel{?}{=} - \frac{1}{16\pi^2}
 h^4 \phi^4 \ln \frac{h^2 \phi^2}{k^2} + \mathcal{O}\left( \frac{ h^2\phi^2}{\Lambda^2} \right).
\label{eq:MFULinfty}
\end{equation}
This in fact looks like an unstable interaction potential for large
$\phi$ (in obvious contradiction with \eqref{eq:MFU}). This corollary
of the original puzzle is resolved by noting that the instability sets
in for large fields where $\left( \frac{ h^2\phi^2}{\Lambda^2} \right)
\sim \mathcal{O}(1)$, i.e., where the large-$\Lambda$ expansion is no
longer justified. We emphasize that our line of argument does not rely
on the momentum-cutoff regularization, but is identical for
gauge-invariant regulators such as propertime or zeta function
regularization. Some care is required for dimensional regularization
which fails to keep track of the explicit cutoff dependence, since it
is a projection (on log divergencies) rather than a regularization
scheme. For details, see \cite{Gies:2014xha}.

To summarize: In order to get global information about the stability
of the Higgs effective potential in the standard model, it is
advisable to \textit{(i)} keep track of the cutoff $\Lambda$ or the RG
scale explicitly, and \textit{(ii)} study the features of the
potential globally. Both aspects can be taken care of with the
functional RG.

\section{Higgs mass bounds as a UV to IR mapping}

Since we keep the cutoff finite in our analysis, we can also use it as
our explicit renormalization scale where we fix all parameters of the
theory in the form of specifying the microscopic action
$S_\Lambda$. As there is no direct experimental information available about
the parameters in this action, we may start from a generic action
\begin{align}
  S_\Lambda &= S_\Lambda(\mL, \lL, \lambda_{3,\Lambda}, \dots, \hL, \dots)  \notag \\
  &= \dots + \frac{1}{2} \mL \phi^2+ \frac{1}{8} \lL \phi^4 + \frac{1}{48} \frac{\lambda_{3,\Lambda}}{\Lambda^2} \phi^6+ \dots + i\hL \phi \bar{\psi}\psi + \dots
\end{align}
that includes higher-order operators such as $\sim \lambda_{3,\Lambda}
\phi^6$, etc. On the level of the bare action, they are not forbidden
at all. In fact, phenomenological studies involving such operators are
common in this context
\cite{Datta:1996ni,Barbieri:1999tm,Grzadkowski:2001vb,Burgess:2001tj,Barger:2003rs,Blum:2015rpa}. Wilson's
powercounting arguments of renormalization tell us that these
higher-order operators die out rapidly toward the IR and thus do not
exert a sizable influence on the long-range observables (as long as
the flow is dominated by the weak-coupling Gau\ss{}ian fixed point
regime). Still, these operators can exert an influence on the IR flow
itself. The renormalization flow, i.e., averaging over the
fluctuations of all quantum fields, now provides a mapping of the bare
action onto the renormalized effective action $S_\Lambda \to
\Gamma$. The latter encodes the dynamics of the theory in the IR and
thus can be more directly parametrized in terms of the long-range
observables. Any measured quantity therefore imposes a constraint on
the form of $\Gamma$ and thus indirectly on the form of $S_\Lambda$.

In the present work, we use the expectation value of the Higgs
field $v\simeq 246$GeV and the top quark mass $\mtop\simeq 173$GeV as
input. This fixes two parameters of the bare action $S_\Lambda$,
e.g. $\mL$ and $\hL$, but leaves the mass of the Higgs boson as a
function of all other parameters of the bare action and of the cutoff
itself, $\mH=\mH[S_\Lambda;\Lambda]$. The underlying mapping
$S_\Lambda \to \Gamma$ can be approximated in various ways,
e.g. perturbatively in the weak-coupling regime. Most useful in this
regime are also a mean-field or extended-mean-field approximation
which lead to fully analytical results, see \cite{Gies:2013fua}. A
more comprehensive picture is obtained with the functional RG,
yielding trustworthy results also at stronger coupling.

The functional RG can be formulated in terms of a flow equation for
the effective action $\Gamma_k$ interpolating between the bare action
$\Gamma_{k=\Lambda}=S_{\Lambda}$ and the 1PI effective action
$\Gamma=\Gamma_{k=0}$. This flow is obtained as the solution to the Wetterich equation
\cite{Wetterich:1992yh},
\begin{equation}
\pat \Gamma_k= \frac{1}{2} \Tr \Big[ (\Gamma_k^{(2)}+R_k)^{-1} \pat R_k \Big],\quad \pat = k\frac{d}{dk},
\label{eq:Wetterich}
\end{equation}
where $R_k$ is a regulator specifying the details of the Wilsonian
regularization near the momentum shell $p\simeq k$, and
$\Gamma_k^{(2)}+R_k$ is the full inverse propagator at scale $k$, for
details see
\cite{Berges:2000ew,Pawlowski:2005xe,Gies:2006wv,Delamotte:2007pf,Braun:2011pp}.
We solve the flow in the space of actions that can be parametrized by 
\begin{equation}
\Gamma_k = \int d^4 x \left( \frac{Z_{\phi,k}}{2} (\partial_\mu \phi)^2 + U_k(\phi) + Z_{\psi,k} \bar{\psi} i \slashed{\partial} \psi + i h_k \phi\bar{\psi} \psi \right),
\end{equation}
where the wave function renormalizations $Z_{\phi/\psi,k}$, the full
potential $U_k$, and the Yukawa coupling $h_k$ flow with the
RG. Solving the flow with $S_\Lambda$ as UV boundary condition (and
fixing the expectation value $v$ and the top mass $\mtop$), we can
read off the Higgs mass as the curvature of the (renormalized)
potential at the minimum $\mH^2=U_{k\to 0}''(v)/Z_{\phi,k\to 0}$.

\begin{figure}[t]
\includegraphics[width=0.48\columnwidth]{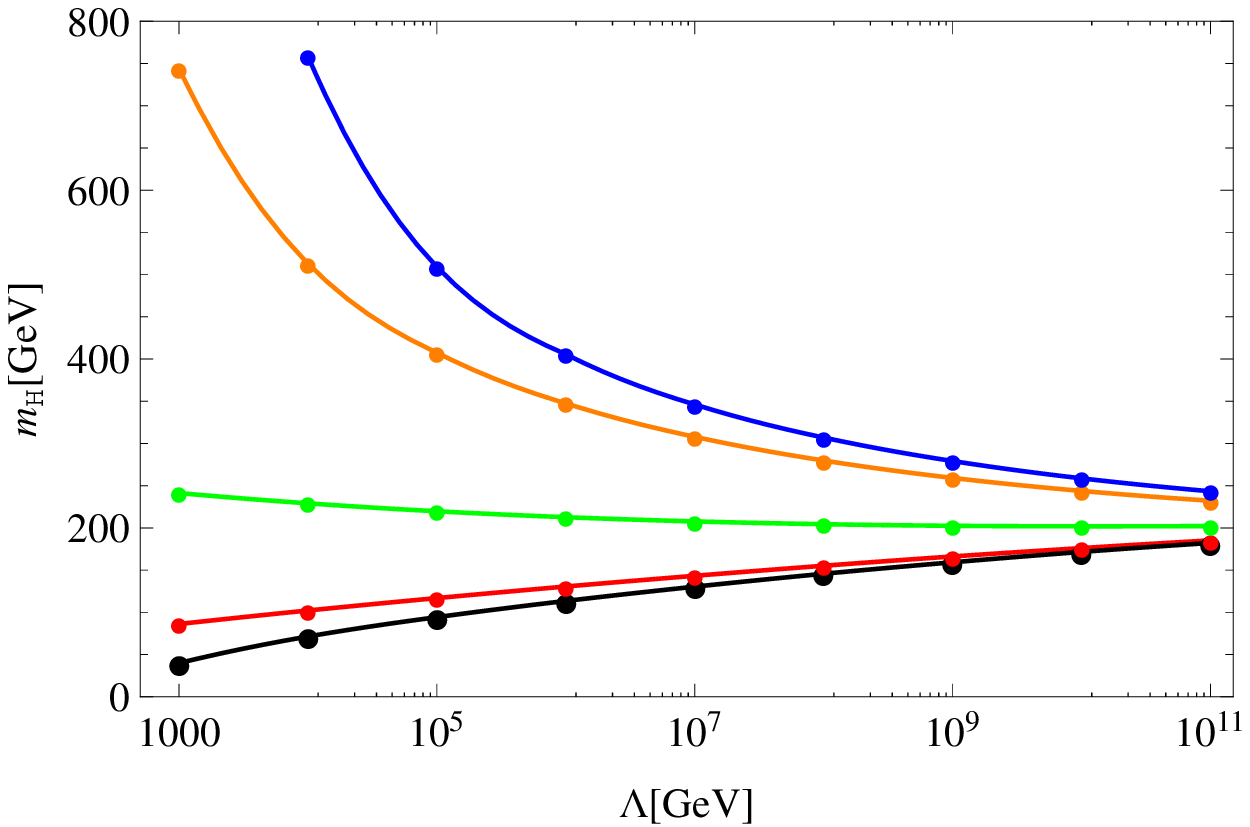}\hfill%
\includegraphics[width=0.48\columnwidth]{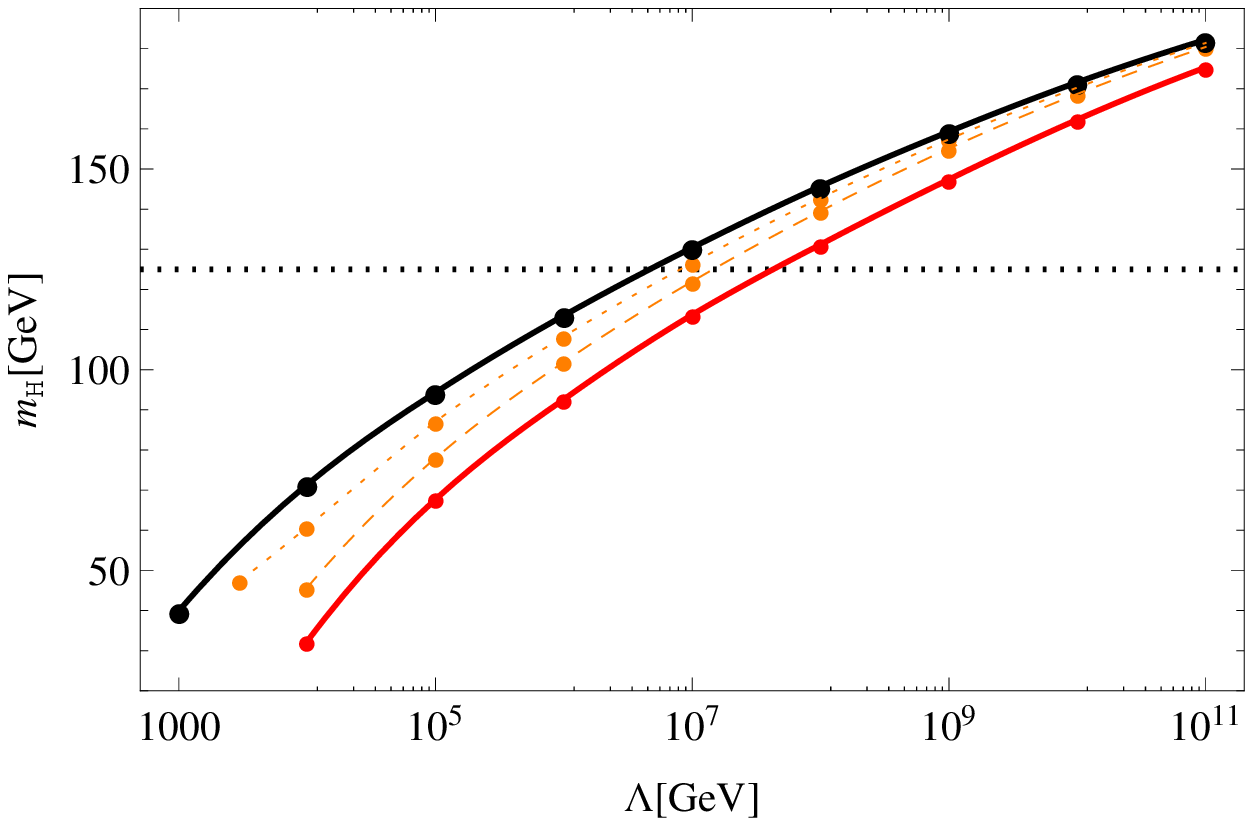}
\caption{Left: Higgs mass $\mH(\lL;\Lambda)$ versus cutoff $\Lambda$
  for a $\phi^4$-type bare potential for $\lL=0$ (black, lower bound)
  and $\lL=0.1,1,10,100$ from bottom to top (red to blue). Right:
  ``lower bound'' with a $\phi^4$-type bare potential (black) in
  comparison with Higgs masses obtained with an inclusion of a bare
  $\lambda_{3,\Lambda}\phi^6$ operator (red line); intermediate orange
  dashed lines correspond to $\lambda_{3,\Lambda}=3$ and
  $\lL=-0.05,-0.08$. The ``lower bound'' can clearly be relaxed; the
  data agrees with \cite{Gies:2013fua}.}
\label{fig:1}
\end{figure}

Restricting the bare actions to the ``renormalizable operators'' of
\Eqref{eq:action}, i.e., $U_\Lambda=\frac{1}{2}\mL \phi^2 +
\frac{1}{8} \lL \phi^4$, the only remaining parameter after fixing $v$
and $\mtop$ is the $\phi^4$ coupling, such that $\mH=\mH(\lL;
\Lambda)$. The resulting Higgs masses in this model are shown in
Fig.~\ref{fig:1} (left panel) as a function of the cutoff $\Lambda$
for various couplings $\lL$ ranging from zero to the strong coupling
region. Though $\mH$ increases monotonically with $\lL$, we observe
that a finite IR window of possible Higgs masses emerges as a result
of the flow. In the present toy model, the center of this window lies
in the region $\simeq 200 \dots 250$GeV. The lower bound is given by
$\lL=0$. Our data indicates that the resulting Higgs mass also
asymptotically approaches an upper Higgs mass bound for increasing
couplings $\lL$ for larger cutoffs $\Lambda$. In this class of bare
$\phi^4$ potentials, the lower bound $\lL=0$ is
dictated by the existence of a well defined functional integral and
thus corresponds to the criterion of absolute stability in this class.

However, the restriction to $\phi^4$ potentials is arbitrary; neither
formal renormalizability arguments nor experimental data serve to
justify such a limitation. As a simple generalization, let us consider
the next higher-order operator in the potential $\sim
\lambda_{3,\Lambda} \phi^6$. In fact, many further operators can be
(and have been) studied, see, e.g.,
\cite{Jakovac:2015kka,Jakovac:2015iqa} for higher-order fermionic
operators or \cite{Gies:2017zwf} for mixed operators. The present
simple $\phi^6$ example suffices to illustrate an important point
here. The stability criterion $\lL\geq0$ of the $\phi^4$ potentials
can, of course, be alleviated by an inclusion of a coupling
$\lambda_{3,\Lambda}>0$. In particular, we can start with a negative
$\lL$ in the UV which turns into a positive $\lambda_{2,k}$ at lower
scales by the RG flow, while $\lambda_{3,k}$ becomes small according
to power counting. We observe that we obtain potentials which are
fully stable on all scales but lead to lower Higgs masses than the
``lower bound'' obtained with $\phi^4$ potentials, see
Fig.~\ref{fig:1} (right panel).  This demonstrates that the
conventional lower bound with $\phi^4$ bare potentials can be relaxed
upon the inclusion of higher-dimensional operators at the cutoff
without losing absolute stability.

Our example suggests that the conventional lower bound should be
replaced by a \textit{consistency} bound defined by the smallest
possible value for the Higgs mass as a function of the cutoff
$\Lambda$, derived from the set of all possible microscopic bare
actions $S_\Lambda$ which define a consistent functional integral and
are compatible with an absolutely stable Fermi minimum. Of course,
computing the bound is a complicated minimization problem in an
infinite dimensional space of bare actions with nontrivial
constraints. The lower Higgs masses (red line) in Fig.~\ref{fig:1}
(right panel) thus represent merely a simple example that this
consistency bound is below the conventional lower bound. 

We also observe in this figure that our red-line example appears to
approach the conventional lower bound (black line) for increasing
cutoff values $\Lambda$. This is natural for flows in the vicinity of
the weak-coupling Gau\ss{}ian fixed point: here, the higher-order
operators are power-law depleted by the RG flow. Typically at scales
of $k\sim 10^{-1\dots 3}\Lambda$, the higher-dimensional operators
do no longer contribute significantly to the flow which thus runs
essentially close to that of the $\phi^4$-class towards the IR. Hence,
the possible shift of the bound in the Higgs mass $\Delta \mH$ also is
a decreasing function of $\Lambda$ under the assumption of weak
coupling.

\section{Towards the standard model}

We have verified that the mechanism described above that leads to
concistency bounds below the conventional stability bounds of the
Higgs boson mass is also active in other models sharing further
similarities with the standard model. The typical chiral structure of
the standard model can, for instance, be tested with a correspondingly
chiral model with a Yukawa sector coupling a complex SU(2) scalar doublet
$\phi$ to a left-handed fermion top-bottom doublet
$\psi_{\text{L}}=(t_{\text{L}},b_{\text{L}})$ and the corresponding right handed singlets $t_{\text{R}}$, $b_{\text{R}}$,
\begin{equation}
S_{\Lambda,\text{Yuk}}= \int d^4 x \Big[ i h_{\text{b},\Lambda} (\bar{\psi}_{\text{L}} \phi b_{\text{R}} + \bar{b}_{\text{R}} \phi^\dagger \psi_{\text{L}} ) + i h_{\text{t},\Lambda} (\bar{\psi}_{\text{L}} \phi_{\mathcal{C}} t_{\text{R}} + \bar{t}_{\text{R}} \phi^\dagger_{\mathcal{C}} \psi_{\text{L}} ) \Big],
\end{equation}
featuring the chiral SU(2) symmetry and giving room for two different
Yukawa couplings $h_{\text{t}}$ and $h_{\text{b}}$; here
$\phi_{\mathcal{C}}=i\sigma_2 \phi^\ast$ is the charge conjugated
scalar. The measured mass of the bottom quark $m_{\text{b}}\simeq
4.2$GeV is used to implicitly fix the UV initial condition for the
bottom Yukawa coupling at the cutoff scale $h_{\text{b},\Lambda}$.

\begin{figure}[t]
\includegraphics[width=0.48\columnwidth]{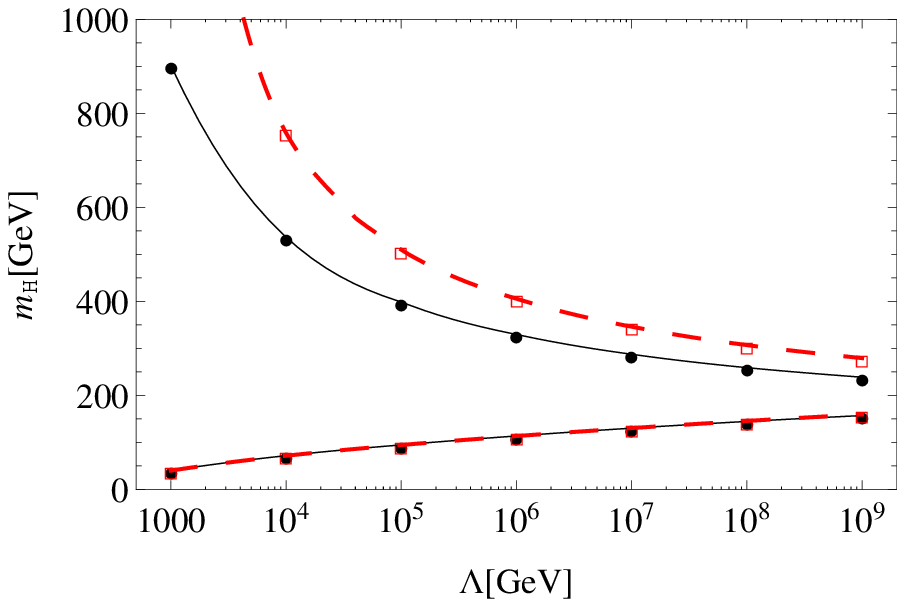}\hfill%
\includegraphics[width=0.48\columnwidth]{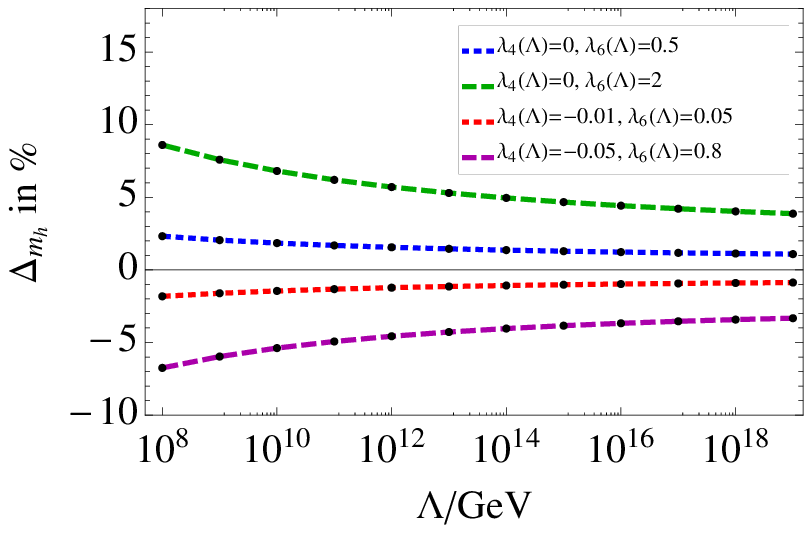}
\caption{Left: Higgs mass $\mH(\lL;\Lambda)$ versus cutoff $\Lambda$
  for a $\phi^4$-type bare potential for $\lL=0$ and $\lL=100$ (bottom
  to top) for the chiral Yukawa model (solid black line) in comparison
  with the $\mathbb{Z}_2$ model (red dashed line); taken from
  \cite{Gies:2014xha}. Right: relative mass shift of the Higgs boson
  mass compared to the conventional lower bound versus cutoff
  $\Lambda$ upon the inclusion of higher-dimensional operators. The
  plot is based on a hybrid model introduced in
  \cite{Eichhorn:2015kea} which quantitatively mimics the standard
  model flow in the Higgs sector. The solid black zero line hence
  corresponds to $\mH\simeq129$ GeV at the Planck scale, and to
  $\mH\simeq125$ GeV near $\Lambda\simeq10^{10}$ GeV. The lowest curve
  accommodates the difference between the conventional lower bound and
  the measured Higgs mass even with Planck-scale operators, but
  corresponds to a metastable Fermi minimum. Higgs mases below the
  conventional lower bounds ($\Delta \mH<0$) and a fully stable Fermi
  minimum can be found by our mechanism with suitable
  higher-dimensional scale terms in the bare action (red short-dashed
  line); taken from \cite{Eichhorn:2015kea} with conventions
  $\lambda_{6}(\Lambda)=\lambda_{3,\Lambda}/6$.}
\label{fig:2}
\end{figure}

It is instructive, to first study the IR window, i.e., the range of
accessible values for the Higgs mass within the class of bare $\phi^4$
potentials \cite{Gies:2014xha}. The result is shown in
Fig.~\ref{fig:2} (left panel) (black solid lines for $\lL=0$ and
$\lL=100$) and compared to the $\mathbb{Z}_2$-symmetric model (red
dashed lines) studied before. We observe that the conventional lower
bound of the two models are almost identical. This confirms the
usefulness of the simple $\mathbb{Z}_2$ Yukawa model for the purpose
of studying the lower bound. The reason for this agreement simply lies
in the fact that the bottom Yukawa coupling is much smaller than that
of the top quark, with the latter dominating the dynamics near the
lower bound. The situation is different at large Higgs self-coupling:
here, the different number of scalar degrees of freedom plays a
substantial role which leads to a narrowing of the IR window, now
having its center near $200$GeV.

This model is also of interest, because it is used for nonperturbative
studies of the Higgs mass bounds in lattice simulations
\cite{Fodor:2007fn,Gerhold:2007yb,Gerhold:2007gx,Gerhold:2009ub}. The
chiral gauge structure of the standard model is so far not accessible
on the lattice, but this chiral model suffices to address quantitative
questions, while extrapolating to the full standard model with the aid
of perturbation theory. In fact, the same mechanism for relaxing the
conventional lower bound by means of higher-order operators that we
discovered for the $\mathbb{Z}_2$ model is also active in this chiral
model \cite{Gies:2013fua,Gies:2014xha}; the corresponding plot upon an
inclusion of a $\sim \lambda_{3,\Lambda} (\phi^\dagger\phi)^3$
operator looks identical to Fig.~\ref{fig:1} (right panel), see Fig.~4
in \cite{Gies:2014xha}. The existence of this mechanism is also
confirmed by lattice studies
\cite{Hegde:2013mks,Chu:2015nha,Chu:2015ula,Akerlund:2015fya},
corroborating the functional RG studies.

The chiral model nevertheless has a technical disadvantage, as the
chirally broken (Fermi) phase necessarily goes along with the
appearance of massless Goldstone modes. In the full standard model,
they disappear through the Higgs mechanism, while they have to be
removed by hand in the model studies both on the lattice as well as in
the functional RG studies. This introduces a quantitatively small
degree of model dependence which cannot be removed as a matter of
principle. In order to avoid the full complexity of the standard
model, but nevertheless acquire quantitatively relevant results, a
hybrid model has been constructed in \cite{Eichhorn:2015kea} that
features a simple scalar $\mathbb{Z}_2$ sector, but includes the QCD
gauge group SU(3) under which the fermions are charged, 
\begin{equation}
 S_\Lambda = \int_x \left[ \frac{1}{4} F_{\mu\nu}^aF_{a\mu\nu}+\frac{1}{2} (\partial_\mu \phi)^2 + U_{\Lambda}(\phi) + \bar{\psi}i\slashed{D}[A]\psi + i{\hL} \, \phi\bar{\psi}_{\text{t}}\psi_{\text{t}} \right],
 \label{eq:gaugedaction}
\end{equation}
but only the top quark has a non-negligible Yukawa coupling to the
Higgs scalar. In addition, the influence of the SU(2) and U(1) gauge
group can be modeled on the level of the perturbative beta
functions. We have shown that the running of perturbatively
renormalizable couplings in the relevant top-Higgs sector of this
hybrid model can indeed be mapped onto that of the full standard model
\cite{Eichhorn:2015kea}.

In this gauged model, the mechanisms induced by higher-order operators
affecting the conventional lower Higgs-mass bound work much in the
same way as in the Yukawa models discussed above. Still, one
quantitative difference arises from the fact that the flow of the top
Yukawa coupling receives an important contribution from the gauge
sector, such that it behaves like an asymptotically free coupling in
the region between the Fermi and, say, the Planck scale. Therefore,
also the strong influence of the top quark fluctuations on the running
of the scalar potential is somewhat reduced. In total, this leads to a
flattening out of the lower Higgs mass curves a la Fig.~\ref{fig:1}
(right panel) toward larger cutoffs as well as to an IR window
centered near smaller Higgs masses. For the standard model (as well as
for our hybrid model), the center of the IR window is near $\simeq 130
\dots 150$ GeV. The conventional lower bound of the Higgs mass has
been under intense investigation in recent years. The precise value,
say for $\Lambda$ equals the Planck mass, depends on the value of the
Yukawa coupling (and to some degree on the value of the strong
coupling constant and heavy gauge boson masses)
\cite{Buttazzo:2013uya,Espinosa:2015kwx,EspinosaThisWorkshop}. Translating the measured
value of the top mass straightforwardly into the Yukawa coupling leads
to a conventional lower bound for absolute stability of $\simeq
129$GeV \cite{Buttazzo:2013uya,Bednyakov:2015sca} which demonstrates the tension
with the measured value near $125$GeV. One important open question is
that of the true quantitative relation between the top mass extracted
from collider data and the corresponding top Yukawa coupling which
appears to be inflicted by intrinsic uncertainties
\cite{Bezrukov:2014ina,Alekhin:2012py}.

In order to study whether the lower-lying consistency bound permits
for stability of the Higgs potential even for the measured low mass of
the Higgs boson, we have determined the influence of the higher-order
operators $\sim\lambda_{3,\Lambda} \phi^6, \dots$ of the scalar
potential onto the lower bound \cite{Eichhorn:2015kea}. As a
quantitative measure, we have extracted the shift of the Higgs boson
mass relative to the conventional lower bound as a function of $\lL$
and $\lambda_{3,\Lambda}$. This shift is displayed in Fig.~\ref{fig:2}
(right panel) for various initial conditions over a wide range of
cutoffs up to the Planck scale. The red short-dashed line shows an
example with a negative value for $\lL$, but a stabilizing positive
$\lambda_{3,\Lambda}$ arranged such that the effective potential stays
stable over all scales. With this example using the model of
\Eqref{eq:gaugedaction} or its hybrid version of
\cite{Eichhorn:2015kea} that exhibits all relevant standard model
features, we can reach Higgs masses $\sim 1\%$, i.e., $\simeq1$GeV
below the conventional lower bound for a cutoff at the Planck
scale. Staying during the full flow within both the Gau\ss{}ian weak
coupling fixed point regime as well as within the class of a stable
Fermi minimum, this $\mathcal{O}(1\%)$ mass shift appears to be a
generic relaxation of the conventional lower bound inducable by Planck
scale operators.

Whether initial conditions away from the Gau\ss{}ian weak coupling
fixed point can lead to a controlled flow with much lower Higgs masses
is an open and equally interesting question. A mass shift of the order
of $5\%$ is straightforwardly possible, if a potential with a
metastable (but possibly long-lived) Fermi minimum is acceptable. An
example is given by parameters leading to the long-dashed purple line in
Fig.~\ref{fig:2} (right panel), cf. the discussion in the next
section. The further two examples in Fig.~\ref{fig:2} (right panel)
show that non-vanishing Planck scale higher-order operators with
$\lL=0$ but $\lambda_{3,\Lambda}>0$ can also induce a positive mass
shift and thus would possibly increase the tension between the
observed value of the Higgs boson mass and Planck scale UV completions
leading to such a bare action.

\section{Higgs potentials with metastable Fermi vacuum}

Our results so far proof that the details of the microscopic bare
action, e.g., Planck scale operators, can take a significant and
quantifiable influence on the lower mass bound for the Higgs boson --
even if we insist on absolute stability of the Fermi vacuum. Within
the conventional perturbative picture, the standard model appears to
be below the conventional lower bound in a meta-stable but
sufficiently long-lived region. This means that the single-scale
potential exhibits another global minimum at large fields in addition
to the Fermi minimum. 

Though our results clearly shift the boundary between the stable and
the metastable parameter region, they, of course, do not exclude the
possibility of a metastable potential. Still, there seems to be a
puzzle: in \Eqref{eq:MFU}, we have shown that the interaction part of
the fermion determinant contributes strictly positively to the scalar
potential. Since the top-fluctuations dominate near the lower bound,
how can a second deeper minimum be generated?

\begin{figure}[t]
\includegraphics[width=0.48\columnwidth]{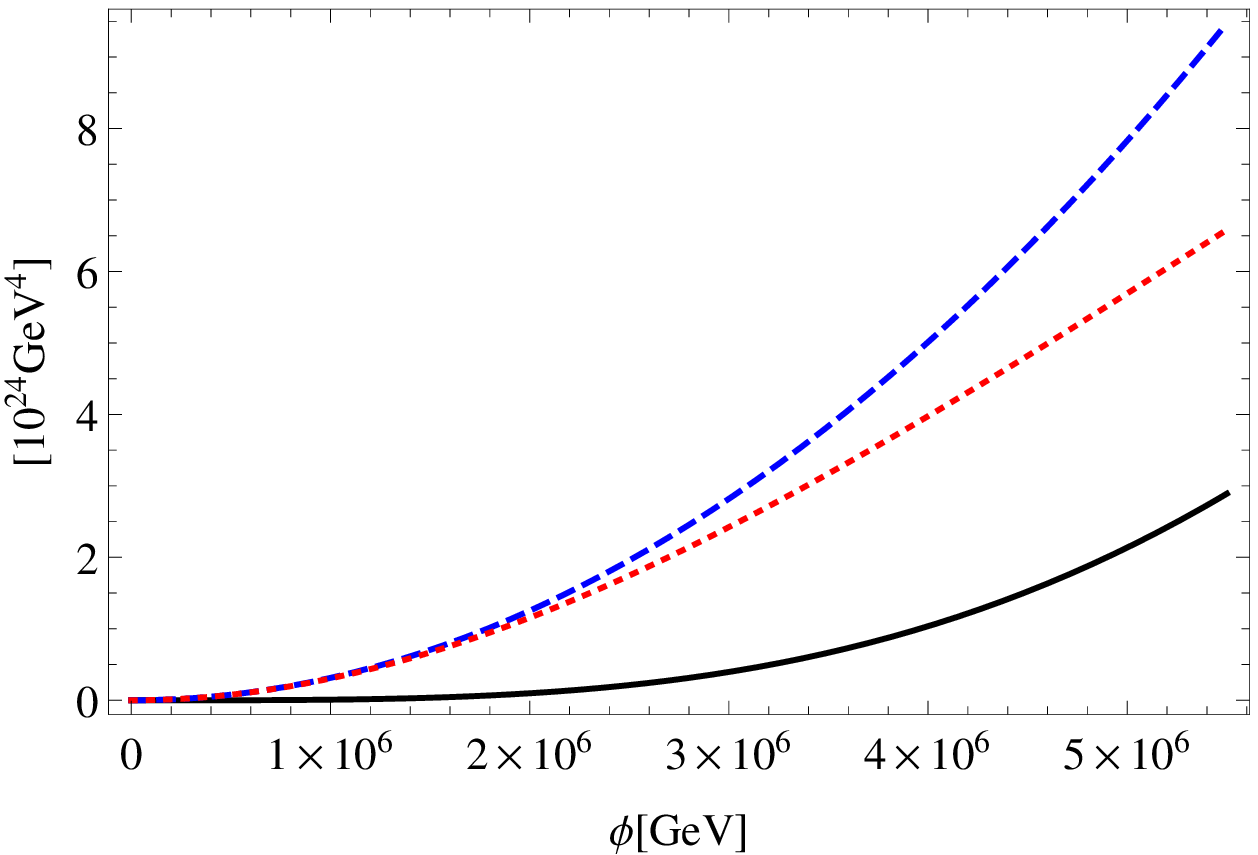}\hfill%
\includegraphics[width=0.48\columnwidth]{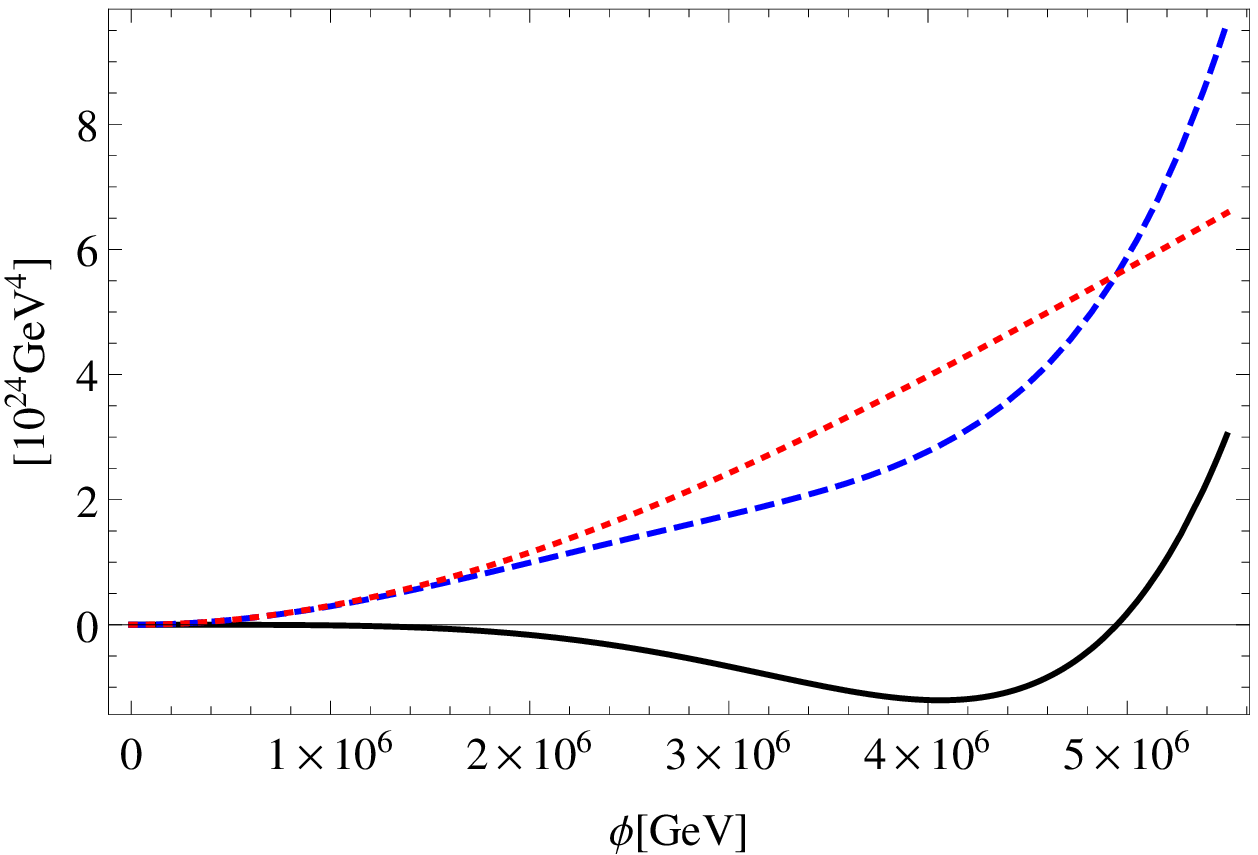}
\caption{Effective mean-field potential (solid line) arising from the
  bare potential (blue long-dashed line) and the top-quark
  fluctuations (minus the fermion determinant) (red short-dashed). The
  effective potential has a minimum at the Fermi scale, $v=246$GeV
  (hardly visible on this scale). Left: initial conditions in the
  class of $\phi^4$-potentials, $\lL=0$,
  $\lambda_{3,\Lambda}=0$. Right: metastability seeded by the bare
  action with $\lL=-0.15$, $\lambda_{3,\Lambda}=3$,
  $\Lambda=10^7$GeV;  taken from \cite{Borchardt:2016xju}.}
\label{fig:3}
\end{figure}

In fact, for microscopic actions with $\phi^4$ bare potentials, this
is not possible provided we start from a well-defined functional
integral requiring $\lL\geq0$. This is illustrated at the mean-field
level in Fig.~\ref{fig:3} (left panel) for our simple Yukawa model,
where the blue long-dashed line denotes the bare potenial with a mass
parameter fixed such that the effective potential (solid black line)
has a proper single Fermi minimum at $v\simeq 246$GeV (not well
visible on this scale), and $\lL,\lambda_{3,\Lambda},\dots=0$. The red
short-dashed line denotes (minus) the fermion determinant which at
mean-field level needs to be subtracted from the bare potential to
yield the effective potential. The resulting effective potential is
globally stable with one minimum at the Fermi scale
\cite{Borchardt:2016xju}.

By contrast, the right panel of Fig.~\ref{fig:3} starts from a bare
action with parameters adjusted such that the bare potential exhibits
a kink (at $\phi\sim 4\times 10^6$GeV in this example). This bare
potential still has only one minimum at $\phi=0$, but in combination
with the contributions from the top-fluctuations (fermion
determinant), the effective potential exhibits two minima, one near the
kink position seeded in the bare action and the second one being the
Fermi minimum by construction (again not well visible on this scale).
We conclude that metastabilities can, of course, occur in the Higgs
sector also in the class of consistent bare actions
\cite{Borchardt:2016xju}. The important point is that this
metastability of the Fermi minimum by a second deeper minimum at large
fields has to be seeded by corresponding structures in the bare
action. The latter would have to be provided by the underlying theory.

While the mean-field approximation is capable of illustrating these
stability/meta-stability features rather well, it does not establish
convexity as is obvious from Fig.~\ref{fig:3} (right). In fact, from
the definition of the effective action as a Legendre transform,
convexity is a built-in property of the effective potential, which is
respected neither at mean-field level nor in conventional perturbation
theory. It is thus a natural question as to whether the convexity
property of the potential has any decisive influence on the picture
developed so far?

As a powerful feature, the structure of the Wetterich equation \eqref{eq:Wetterich}
does establish convexity of the potential for its global solutions
\cite{Litim:2006nn}. We have analyzed this approach to convexity
quantitatively in the present model in \cite{Borchardt:2016xju}. Our
results demonstrate that the convexity generating mechanisms set in in
the deep infrared below the scale where the top quark decouples and the Fermi
minimum forms. For all cases studied in
\cite{Borchardt:2016xju}, there is a clear separation between the
regime where the electroweak mass spectrum arises and the IR regime
where convexity sets in. Hence, we expect convexity to be essentially
irrelevant for the mass spectrum, whereas its influence on
vacuum-decay-rate calculations deserves a more detailed study.

\section{Conclusion}

We have reconsidered the conventional reasoning for the computation of
bounds on the Higgs mass arising from the assumption that the standard
model provides a quantitatively valid description of nature up to some
high-energy scale $\Lambda$. Because of our ignorance of the physics
at that scale $\Lambda$, the micropscopic action $S_\Lambda$ serving
as a UV boundary condition is largely undetermined. As a consequence,
higher-dimensional operators have to be expected to be present at the
high scale. While these operators do not take a direct influence on
long-range observables provided the standard model is close to the
weak-coupling fixed point, they can modify the RG flow near the high
scale, leaving an indirect imprint on low-energy physics. 

We have illustrated these findings with the example of consistent
(effective) field theories that allow for Higgs boson masses below the
conventional lower bound. This mechanism triggered by
higher-dimensional operators is also present in the standard
model. Our estimates based on a variety of model studies suggest that
even Planck scale operators can induce a relaxation of the
conventional lower bound on the order of 1\% for the absolute
stability bound. While we have considered here the influence of only
one $\phi^6$-type operator as an example, similar features occur, for
instance, upon the inclusion of higher-order fermionic operators
\cite{Jakovac:2015kka,Jakovac:2015iqa} or mixed operators
\cite{Gies:2017zwf} as well as in theories with additional scalar
fields, e.g., dark matter candidates \cite{Eichhorn:2014qka}. Most of
the quantitative studies so far have explored only initial conditions
which are already sufficiently close to the Gau\ss{}ian fixed
point. We emphasize that it remains an interesting open question to
study how far the true consistency bounds can be pushed if also
strongly coupled regions are included in the analysis. The functional
RG advocated in this work is ideally suited for this problem.

Since even the conventional estimates are compatible with a metastable
but sufficiently long-lived Higgs vacuum state, it is a legitimate
question as to whether there is any relevant difference for
contemporary phenomenology, depending on which scenario is ultimately
realized. In fact, the interplay between stability, metastability and
cosmology is currently actively studied within various cosmological
models
\cite{Bezrukov:2014ipa,Moss:2015fma,Espinosa:2015qea,Herranen:2015ima,%
  EspinosaThisWorkshop,MarkkanenThisWorkshop}. If such considerations
favored absolute stability but the tension with the observed Higgs
mass and the conventional bounds persisted, this could point to new
physics below the Planck scale (according to the conventional
interpretation) or correspond to a first measurement of properties of
the microscopic action with a sufficiently deep consistency bound.
With respect to cosmology, also the thermal evolution of the Higgs
potential is of substantial interest. In this respect,
higher-dimensional operators can also exert an influence on the nature
of the thermal phase transition, as has already been observed in
lattice simulations \cite{Akerlund:2015gfy}. It is an interesting
question as to whether the set of bare actions contains relevant
microscopic initial conditions for which the electroweak phase
transition is sufficiently strongly first order in order to support
electroweak baryogenesis.

A particularly fascinating scenario would arise, if the Higgs boson
mass happened to lie close to the value that corresponds to an
exactly flat interaction potential. Then the UV theory has to
guarantee such an exceptional matching condition at the high scale. In
fact, asymptotically safe gravity had been suggested for such a
scenario already before the discovery of the Higgs boson
\cite{Shaposhnikov:2009pv}, see also \cite{Eichhorn:2017eht}. Purely
within particle field theory, such a scenario would be most natural in
theories where also the Higgs sector becomes asymptotically free -- a
long-standing idea that currently witnesses new attraction \cite{Holdom:2014hla,Gies:2015lia,Gies:2016kkk,Helmboldt:2016mpi,Hansen:2017pwe}.

%%Implications: Cosmological ... (Rubio, Espinosa, Tommi) ... handel on UV action ... Special point in theory space (asymptotic freedom?)

\enlargethispage{20pt}

%\ethics{Insert ethics statement here if applicable.}

%\dataccess{Insert details of how to access any supporting data here.}

%\aucontribute{For manuscripts with two or more authors, insert details of the authors’ contributions here. This should take the form: 'AB caried out the experiments. CD performed the data analysis. EF conceived of and designed the study, and drafted the manuscript All auhtors read and approved the manuscript'.}

%\competing{Insert any competing interests here. If you have no competing interests please state 'The author(s) declare that they have no competing interests’.}

\funding{This work received funding support by the DFG under Grants
  No. GRK1523/2 and No. Gi328/7-1. R.S.~acknowledges support by the
  Carl-Zeiss foundation.}

\ack{We are grateful to J.~Borchardt, A.~Eichhorn, C.~Gneiting,
  J.~Jaeckel, T.~Plehn, M.M.~Scherer, and M.~Warschinke for
  collaboration on various parts of the topic. We thank A.~Rajantje,
  M.~Fairbairn, T.~Markkanen, and A.~Eichhorn for their efforts in
  organizing this productive and enjoyable workshop.}

%\disclaimer{Insert disclaimer text here if applicable.}

%%%%%%%%%% Insert bibliography here %%%%%%%%%%%%%%

\bibliographystyle{vancouver}
\bibliography{bibliography}

\end{document}